\documentclass[10pt]{article}

\usepackage{cite}

\usepackage{authblk}

\usepackage{graphicx} 
\usepackage[scriptsize,nooneline,hang]{caption}
\usepackage[hang,nooneline,scriptsize]{subfigure}
\usepackage{pstricks,pstricks-add}
\usepackage{pst-plot}
\usepackage{color}
\usepackage{enumerate}

\usepackage{amsmath,amsfonts,bbm,dsfont,mathrsfs} 

\newcommand{\dd}{{\mathrm{d}}}
\newcommand{\mQ}{{\mathcal{Q}}}
\newcommand{\mP}{{\mathcal{P}}}

\begin{document}

\title{Geodesic equations for particles and light in the black spindle spacetime}
\author{Kai Flathmann}
\author{Noa Wassermann}
\affil{Institut f\"ur Physik, Universit\"at Oldenburg, D--26111 Oldenburg, Germany}

\maketitle
\begin{abstract}
In this paper we derive the geodesic equation for massive particles and light for the black spindle spacetime. The solution for light can be formulated in terms of the Weierstra{\ss} $\wp$-, $\sigma$- and $\zeta$-function, whereas a part of the solutions for massive particles is given in terms of derivatives of the Kleinian $\sigma$-function. We analyze the possible orbit types using parametric diagrams and effective potentials. Furthermore we visualize the orbits in a coordinate system, where the spindle-like topology of the horizon is visible.
\end{abstract}

\section{Introduction}
The recent years in gravitational physics were exciting for experimental physics as well as theoretical physics. The first detection of gravitational waves of a binary black hole merger \cite{Abbot:2016} and of a binary neutron star merger \cite{Abbot:2017} by LIGO/VIRGO mark the starting point of the era of multimessenger astronomy. A few years later the Event Horizon Telescope produced the first picture of a black hole shadow \cite{EHT:2019}. A crucial ingredient for the calculation of the shadow of a black hole is sufficient knowledge about the geodesics of light in an axially symmetric spacetime. For example for a spherically symmetric spacetime (e.g. Schwarzschild spacetime) the size of the shadow is given by the size of the innermost circular photon orbit or photon sphere. \\
In four dimensions the event horizon of axisymmetric black holes, that are asymptotically flat obey the dominant energy condition and, therefore, are topologically spheres \cite{Hawking:1972, Hawking:1973}. If one relaxes one or more of these assumptions (e.g. asympotically anti-de Sitter) horizons of black holes can have different topology. One possibility is the Black Spindle spacetime found by Klemm in 2014 \cite{Klemm:2014} as a subclass of the Carter-Pleba\'{n}ski solution \cite{Carter:1968,Plebanski:1975}. The Black Spindle spacetime describes black holes with a noncompact event horizon while still having a finite volume. This results in a finite entropy. Topologically the event horizons are spheres with two punctures and the spacetime represents the ultraspinning limit of the Kerr-Newman-AdS spacetime. This interesting type of solutions was first obtained in
\cite{Gnecchi:2013mja}, and immediately received much interest, in particular, with respect to their thermodynamics (see e.g. \cite{Hennigar:2014cfa,Hennigar:2015cja,Hennigar:2015gan}). Subsequently further types of black holes with noncompact event horizons of finite area were found: black bottles, whose event horizons are topologically spheres with a single puncture \cite{Chen:2016rjt}. \\
To analyze the structure of spacetime one of the most powerful tools is geodesics. The first to solve the geodesic equation for the Schwarzschild black hole was Hagihara \cite{Hagihara:1931}. He solved the equations of motion in terms of the elliptic Weierstra{\ss} $\wp$-function. Elliptic functions have been used to solve the equations of motion for various black hole and wormhole spacetimes \cite{Kagramanova:2010bk,Grunau:2010gd,Kagramanova:2012hw,Hackmann:2013pva,Cebeci:2016,Flathmann:2015,Chatterjee:2019}. When higher order polynomials ($>4$) are involved in the geodesic equation the solutions are based on the Jacobi inversion problem \cite{Hackmann:2008a,Hackmann:2008b}. These solutions already have been used for spacetimes, where the horizons aren't topologically spheres. In \cite{Grunau:2012a} the authors analyzed the geodesics of test particles and light in the Singly Spinning Black Ring Spacetime and in \cite{Grunau:2013a} for the (Charged) Doubly Spinning Black Ring Spacetime. In addition the geodesic motion in the (rotating) black string spacetime have been studied in \cite{Grunau:2013b}.   \\
\newpage
Therefore in this article we study the geodesic motion of test particles and light for the black spindle \cite{Klemm:2014} to analyze the structure of the spacetime. The structure of the article is as follows. In Sec. \ref{sec:spacetime} we review the metric of the Black Spindle spacetime and analyze its properties. Sec. \ref{sec:eom} is devoted to the derivation of the equations of motions and the analysis of possible orbit types can be found in Sec. \ref{sec:class}. The solution of the equations of motion in terms of the Weierstra{\ss} $\wp$-, $\sigma$- and $\zeta$-function and of the Kleinian $\sigma$-function are presented in Sec. \ref{sec:sol}. The visualization of the orbits can be found in Sec. \ref{sec:orbits}. We conclude the article in Sec. \ref{sec:conclusion}.  
\section{The Black Spindle}\label{sec:spacetime}
\begin{figure}[ht]
\centering
\includegraphics[width=0.5\textwidth]{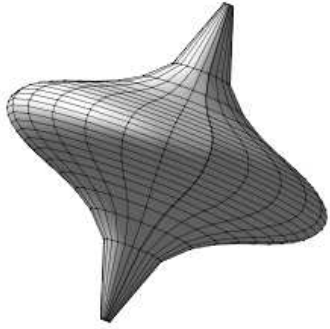}
\caption{Embedding of the horizon in cylindrical coordinates.}
\label{pic:spindle}
\end{figure}

The four dimensional spacetime of a black spindle described by Klemm is given by the metric \cite{Klemm:2014}
\begin{equation}
\dd s^2= -\frac{\mQ}{p^2+q^2}\left(\dd\tau-\dd\sigma\right)^2+\frac{\mP}{p^2+q^2}\left(\dd\tau+\dd\sigma\right)^2+\frac{p^2+q^2}{\mQ}\dd q^2+\frac{p^2+q^2}{\mP}\dd p^2 \,,
\label{eqn:metric}
\end{equation}
where
\begin{align}
\mQ =&\left(l+\frac{q^2}{l}\right)^2+P^2+Q^2-2mq \\
\mP =&\frac{\left(p^2-l^2\right)^2}{l^2}\,.
\label{eqn:metricfunctions}
\end{align}
Here $l^2=-\frac{3}{\Lambda}$ is related to the cosmological constant $\Lambda$, $m$ is the mass parameter and $Q$ and $P$ are the electric and magnetic charge of the black hole. For simplification we define the combined electromagnetic charge parameter $C^2=P^2+Q^2$. 
The condition $q=p=0$ defines the location of the curvature singularity, whereas for $\mQ=0$ we find up to two horizons $q_{\pm}$. To guarantee that the singularity is hidden behind the horizons (e.g real valued horizons), the condition \cite{Klemm:2014,Ling:2015}
\begin{equation}
m \geq 2q_{h,0}\left(\frac{q_{h,0}^2}{l^2}+1\right)\,,
\end{equation}
with
\begin{equation}
q_{h,0}^2=\frac{l^2}{3}\left(-1+\sqrt{4+\frac{3C^2}{l^2}}\right)
\end{equation}
has to be fulfilled. The \"{}$=$\"{} sign corresponds to an extremal black hole, where both horizons coincide. To make the spindle shape of the horizons visible, we embed the spacetime in cylindrical coordinates. By fixing $\tau=0$ and $q=q_{\pm}$, we can transform the metric in Eqn. \ref{eqn:metric} into
\begin{equation}\label{eqn:metric_embedd}
 \dd s^2=\frac{p^2+q_{\pm}^2}{\mP}\dd p^2+\frac{\mP q_{\pm}^4}{p^2+q_{\pm}^2}\dd\sigma^2\,.
\end{equation}
A standard transformation (see for example \cite{Ling:2015}) can be used to obtain the angular coordinate $\phi=-l\sigma$. In this case it is possible to write Eqn. \ref{eqn:metric_embedd} as
\begin{equation}
 \dd s^2=\dd\rho^2+\rho^2\dd\phi^2+\dd z^2 \,,
\end{equation}
with
\begin{equation*}
 \rho^2=\frac{\mP q_{\pm}^4}{l^2\left(p^2+q_{\pm}^2\right)}
\end{equation*}
and
\begin{equation*}
 \left(\frac{\dd z}{\dd p}\right)^2 = \frac{p^2+q_{\pm}^2}{\mP}-\left(\frac{\dd\rho}{\dd p}\right)^2\,.
\end{equation*}
We will solve this differential equation numerically. This procedure is often used for wormhole solutions (e.g. \cite{Willenborg:2018}). See Fig. \ref{pic:spindle} for an example plot.

\section{The geodesic equations}\label{sec:eom}
The Hamilton-Jacobi equation 
\begin{equation}
\frac{\partial S}{\partial \lambda}+\frac{1}{2}g^{\mu\nu}\frac{\partial S}{\partial x^{\mu}}\frac{\partial S}{\partial x^{\nu}}=0
\end{equation}
can be solved with the ansatz
\begin{equation}
S=\frac{1}{2}\delta\lambda-E\tau+L\sigma++S_q(q)+S_p(p) \,.
\label{eqn:action}
\end{equation}
Here $E$ denotes the energy of the particle, $L$ its angular momentum and $\delta$ is either equal to $1$ for massive particles or vanishes for light. For massive particles, $\lambda$ is related to the proper time and for light it is an affine parameter along the geodesics. The Hamilton-Jacobi equation separates with the help of the Carter constant $K$ \cite{Carter:1968rr}. Therefore we derive the equations of motion for each coordinate
\begin{align}
\frac{\dd q}{\dd\gamma} &= \sqrt{X} \nonumber\\
\frac{\dd p}{\dd\gamma} &= \sqrt{Y} \nonumber\\
\frac{\dd\sigma}{\dd\gamma} &= \frac{L-Ep^2}{\mP}-\frac{L-Eq^2}{\mQ}\nonumber\\
\frac{\dd\tau}{\dd\gamma} &= \frac{Lp^2-Ep^4}{\mP}+\frac{Eq^4+Lq^2}{\mQ}\,.
\label{eqn:EOM}
\end{align}
For simplification we have defined the mino-time $\gamma$ \cite{Mino:2003yg} with $\dd\lambda=\left(p^2+q^2\right)\dd\gamma$. The functions $X$ and $Y$ are polynomials of order six in $q$ and $p$
\begin{align}
X(q) &= E^2q^4+2ELq^2-\left(\delta q^2+K\right)\mQ+L^2 \nonumber\\
Y(p) &= -E^2p^4+2ELp^2+\mP\left(K-\delta p^2\right)-L^2 \,.
\label{eqn:XY}
\end{align}
\section{Classification of the geodesics}\label{sec:class}
\subsection{Parametric diagrams}
\begin{figure}[!ht]
	\centering
	\subfigure[$\delta=0$]{
		\includegraphics[width=0.45\textwidth]{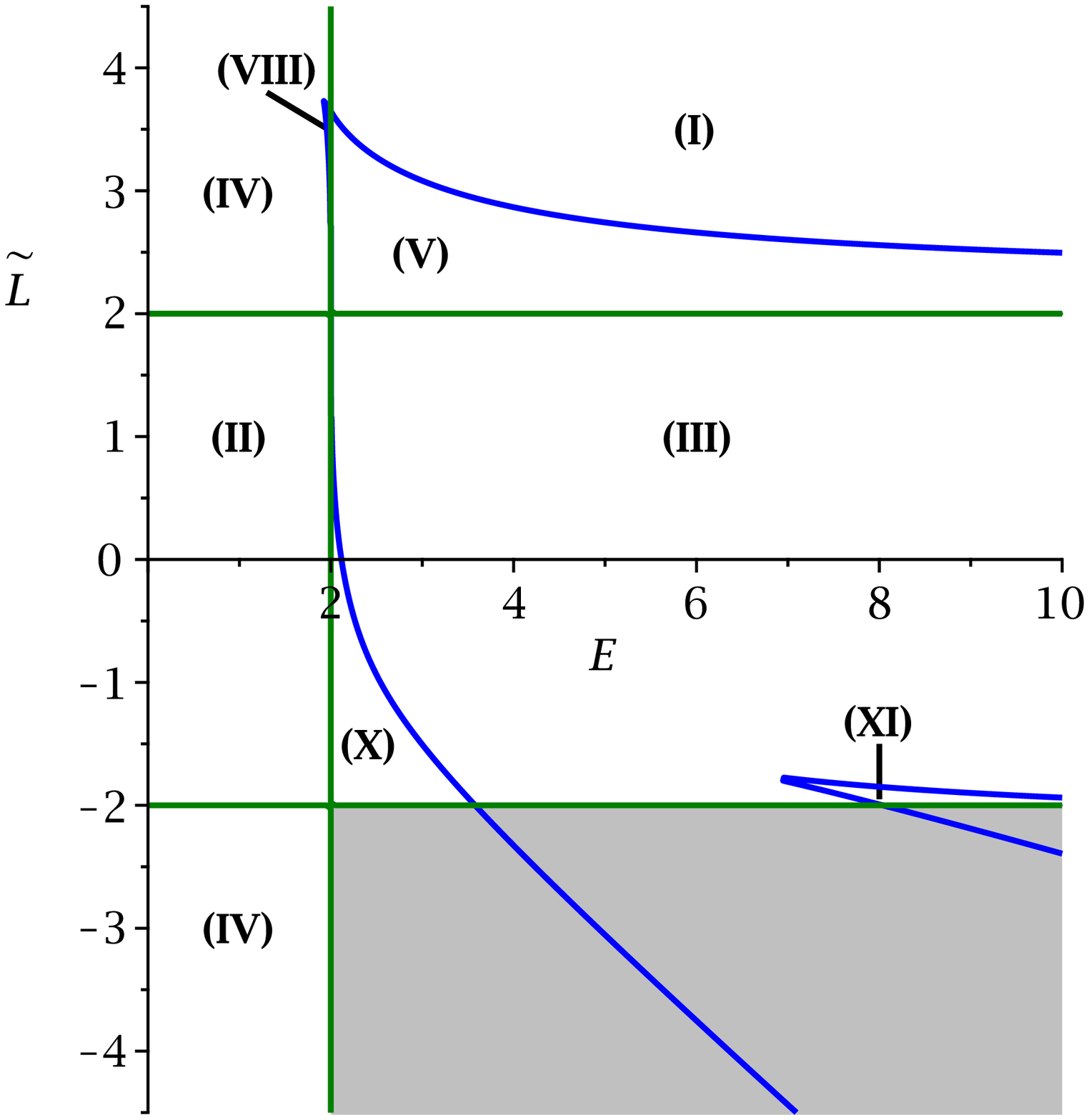}
	}
	\subfigure[$\delta=1$]{
		\includegraphics[width=0.45\textwidth]{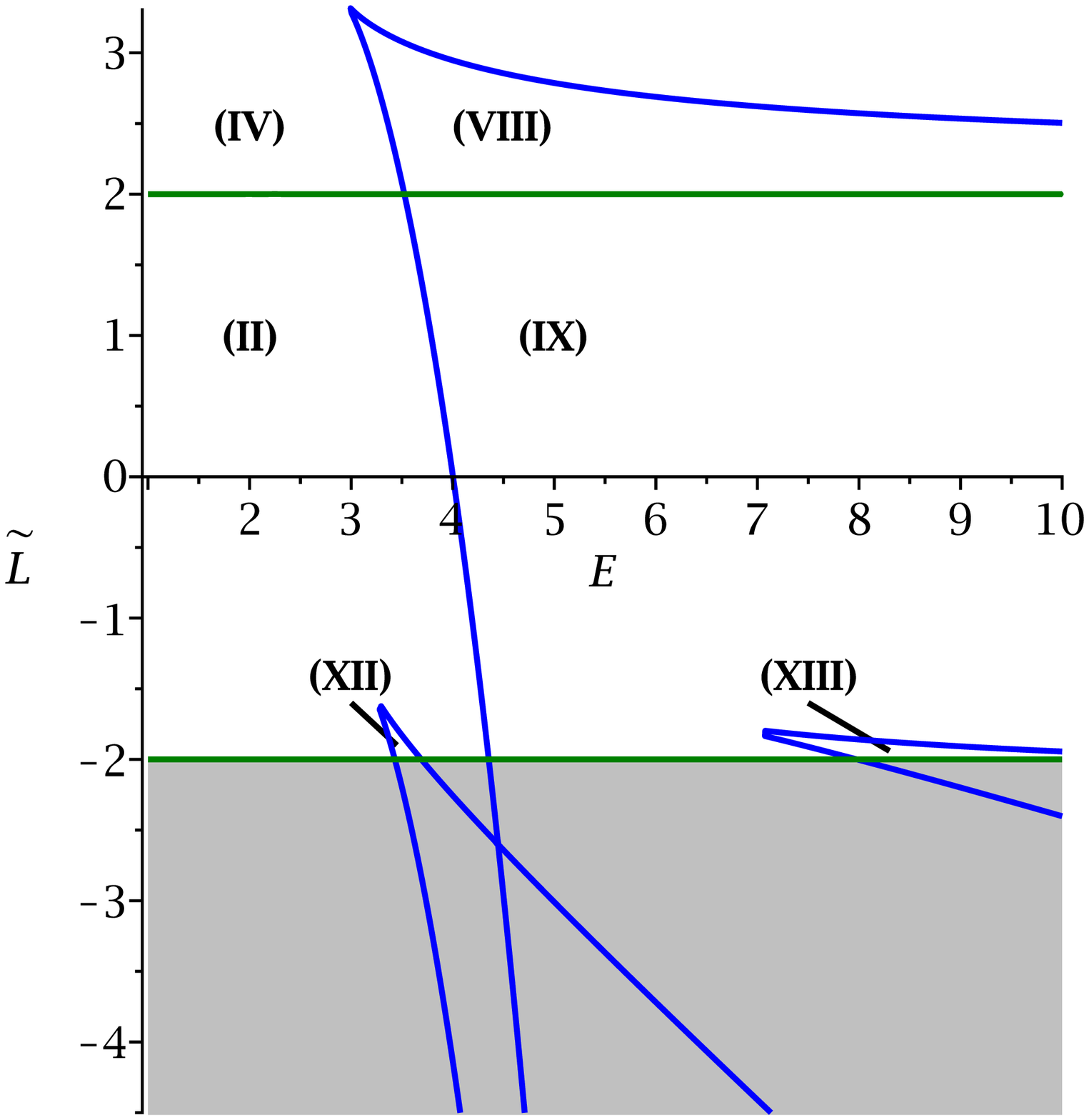}
	}
	\caption{Combined parametric $L-E$ diagrams for the $q$-motion (blue lines) and the $p$-motion (green lines) for $m=m_0+0.1$, $C=0.5$, $K=4$ and $l=1$. In the shaded area there is no geodesic motion possible because $Y(p)<0$ for all $p$.}
 \label{pic:parametricdiagrams}
\end{figure}

To investigate the motion of particles and light in the black spindle spacetime, we have to determine the number of zeros of the polynomials $X$ and $Y$, which correspond to the turning points of the motion. The first tool we are using is parametric diagrams. The number of zeros of $X$ and $Y$ changes if double roots appear. This is the case for
\begin{align}
X(q)&=0\quad \text{and}\quad \frac{\dd X(q)}{\dd q}=0\,, \nonumber \\
Y(p)&=0\quad \text{and}\quad \frac{\dd Y(p)}{\dd p}=0\,.
\end{align}
The combination of these four conditions can be found in Fig. \ref{pic:parametricdiagrams}.
The following regions can be found for the $q$-motion ($q_i<q_{i+1}$)
\begin{enumerate}[(I)]
 \item $X(q)$ has no real zeros and $X(q)>0$. Only Transit orbits (TrO) are possible.
 \item $X(q)$ has two positive zeros $q_1$ and $q_2$ with $X(q)\geq 0$ for $q\in[q_1,q_2]$. Only Many-world bound orbits (MBO) are possible.
 \item $X(q)$ has one positive and one negative zero, where $X(q)\leq 0$ for $q\in[q_1,q_2]$. Possible orbits are Escape orbits (EO) and Two-world escape orbits (TEO).
 \item $X(q)$ has one positive and one negative zero, where $X(q)\geq 0$ for $q\in[q_1,q_2]$. Only Crossover Many-world bound orbit (CMBO) are possible.
 \item $X(q)$ has two negative zeros $q_1$ and $q_2$ with $X(q)\leq 0$ for $q\in[q_1,q_2]$. Possible orbits are EO and Crossover Two-world escape orbits (CTEO).
 \item $X(q)$ has one negative $q_1$ and two positive zeros $q_2,q_3$ with $X(q)\leq 0$ for $q\in[q_1,q_2]$ and $q\in[q_3,q_4]$. Here $q_2\leq q_-$ and $q_3\geq q_+$. EO and MBO are possible.
 \item $X(q)$ has one negative $q_1$ and two positive zeros $q_2,q_3$ with $X(q)\leq 0$ for $q\in[q_1,q_2]$ and $q\in[q_3,q_4]$. Here $q_2,q_3\leq q_-$  EO, BO and TEO are possible.
 \item $X(q)$ has three negative $q_1,q_2,q_3$ and one positive zero $q_4$ with $X(q)\geq 0$ for $q\in[q_1,q_2]$ and $q\in[q_3,q_4]$. Here negative BO and CMBO are possible.
 \item $X(q)$ has two negative $q_1,q_2$ and two positive zeros $q_3,q_4$ with $X(q)\geq 0$ for $q\in[q_1,q_2]$ and $q\in[q_3,q_4]$. Possible orbits are BO and MBO. The BO in this case has always $q\leq 0$.
 \item $X(q)$ has one negative $q_1$ and three positive zeros $q_2,q_3,q_4$ with $X(q)\leq 0$ for $q\in[q_1,q_2]$ and $q\in[q_3,q_4]$. Here $q_3\leq q_-$ and $q_4\geq q_+$. There exist a negative EO an MBO and a positive EO.
 \item $X(q)$ has one negative $q_1$ and three positive zeros $q_2,q_3,q_4$ with $X(q)\leq 0$ for $q\in[q_1,q_2]$ and $q\in[q_3,q_4]$. Here $q_3,q_4\leq q_-$. Possible orbits are EO, BO and TEO.
 \item $X(q)$ has four positive zeros $q_1,q_2,q_3,q_4$ with $X(q)\geq 0$ for $q\in[q_1,q_2]$ and $q\in[q_3,q_4]$. BO and MBO are possible.
 \item $X(q)$ has two negative $q_1,q_2$ and four positive zeros $q_3,q_4,q_5,q_6$ with $X(q)\geq 0$ for $q\in[q_1,q_2]$, $q\in[q_3,q_4]$ and $q\in[q_5,q_6]$. This region is only possible for $\delta=1$. Possible orbits are BO and MBO.
\end{enumerate}
\subsection{Effective potentials} 
\begin{figure}[!ht]
	\centering
	\subfigure[$\delta=0$, $C=0.5$, $l=1$, $m=m_0+0.1$ $K=4$ and $L=3$]{
		\includegraphics[width=0.45\textwidth]{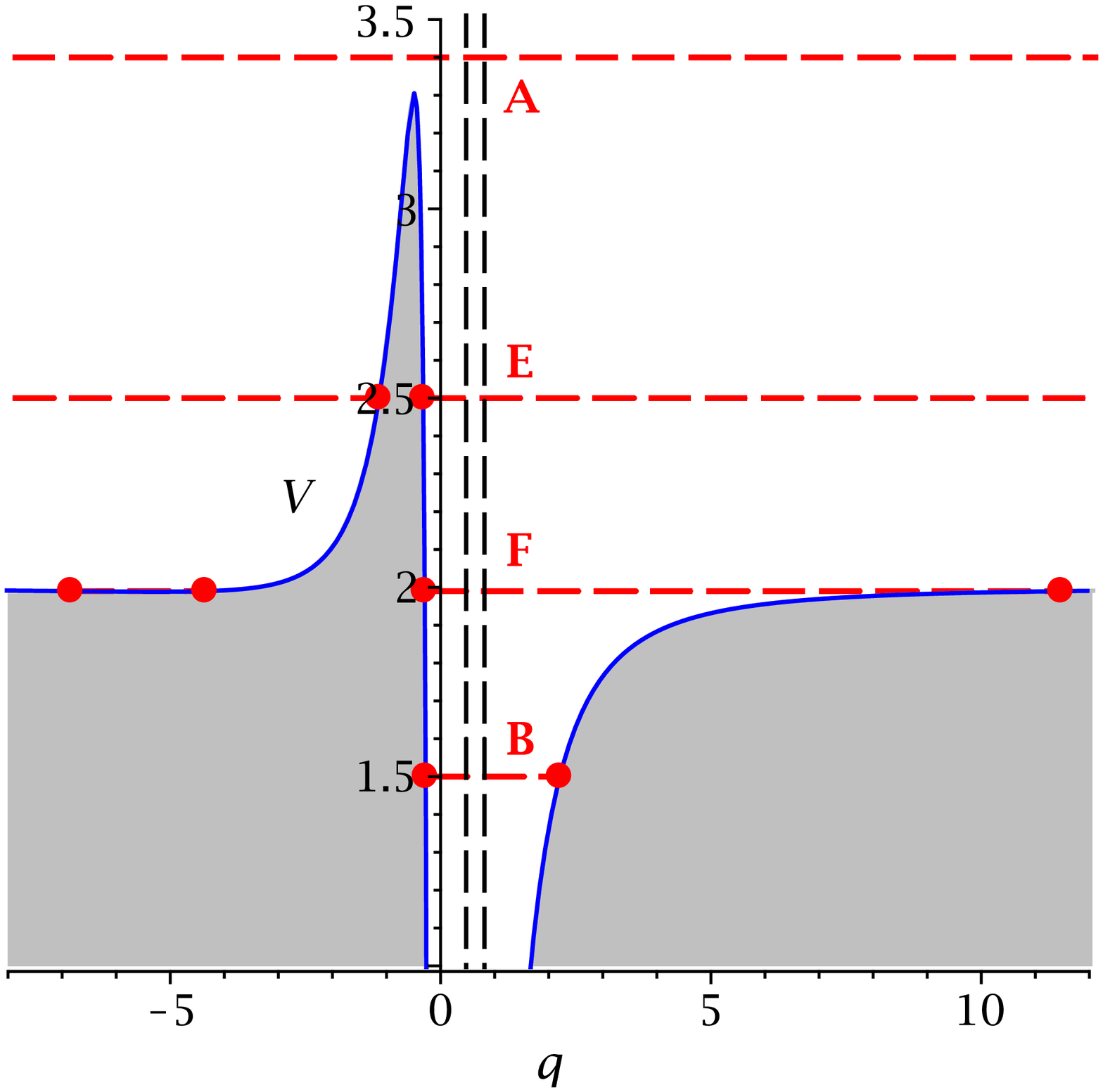}
	}
	\subfigure[$\delta=0$, $C=0.5$, $l=1$, $m=m_0+0.1$ $K=4$ and $L=-1.9$]{
		\includegraphics[width=0.45\textwidth]{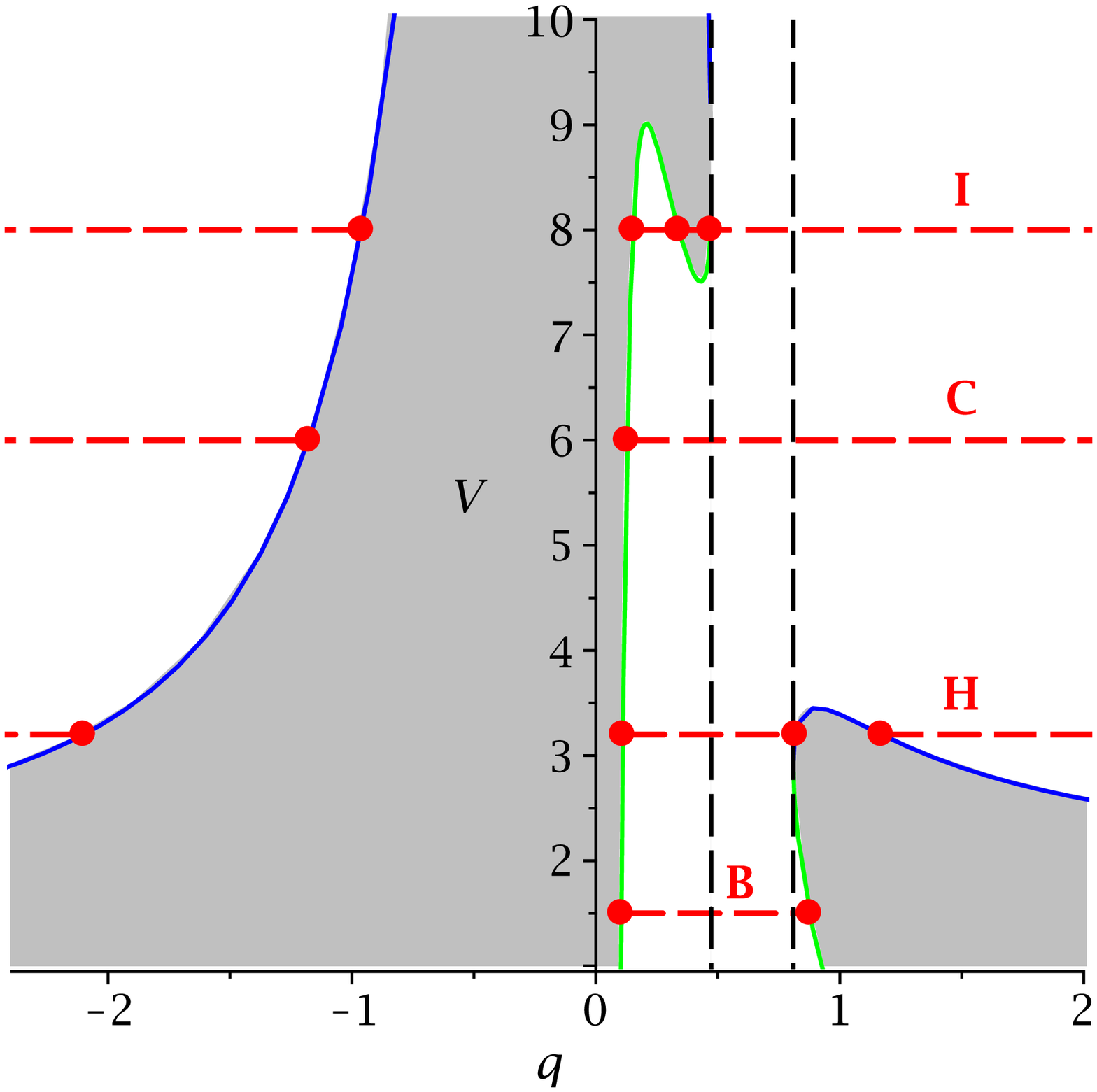}
	}
	\subfigure[$\delta=1$, $C=0.5$, $l=1$, $m=m_0+0.1$ $K=4$ and $L=3$]{
		\includegraphics[width=0.45\textwidth]{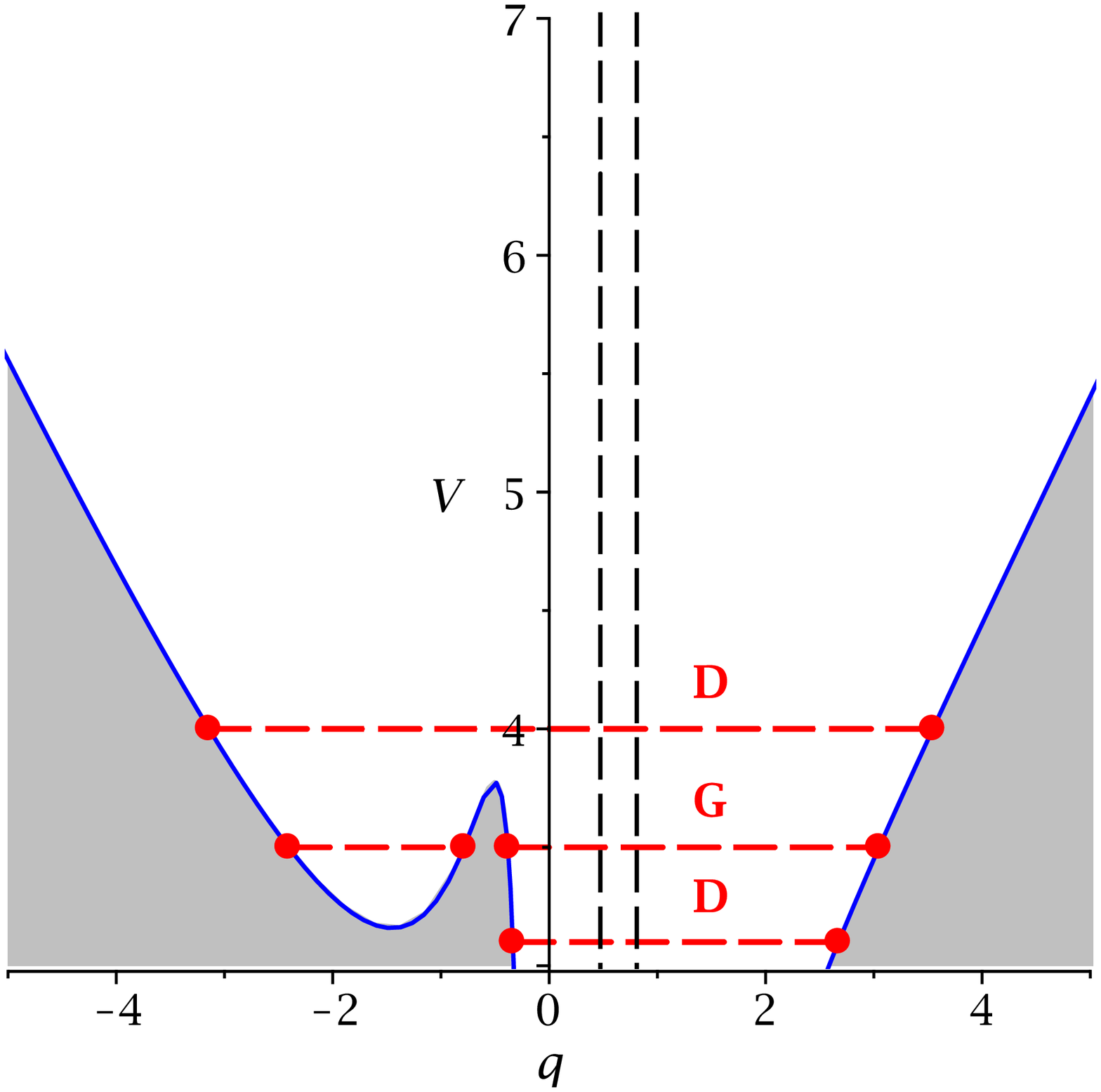}
	}
	\subfigure[$\delta=1$, $C=0.5$, $l=1$, $m=m_0+0.1$ $K=4$ and $L=-1.9$]{
		\includegraphics[width=0.45\textwidth]{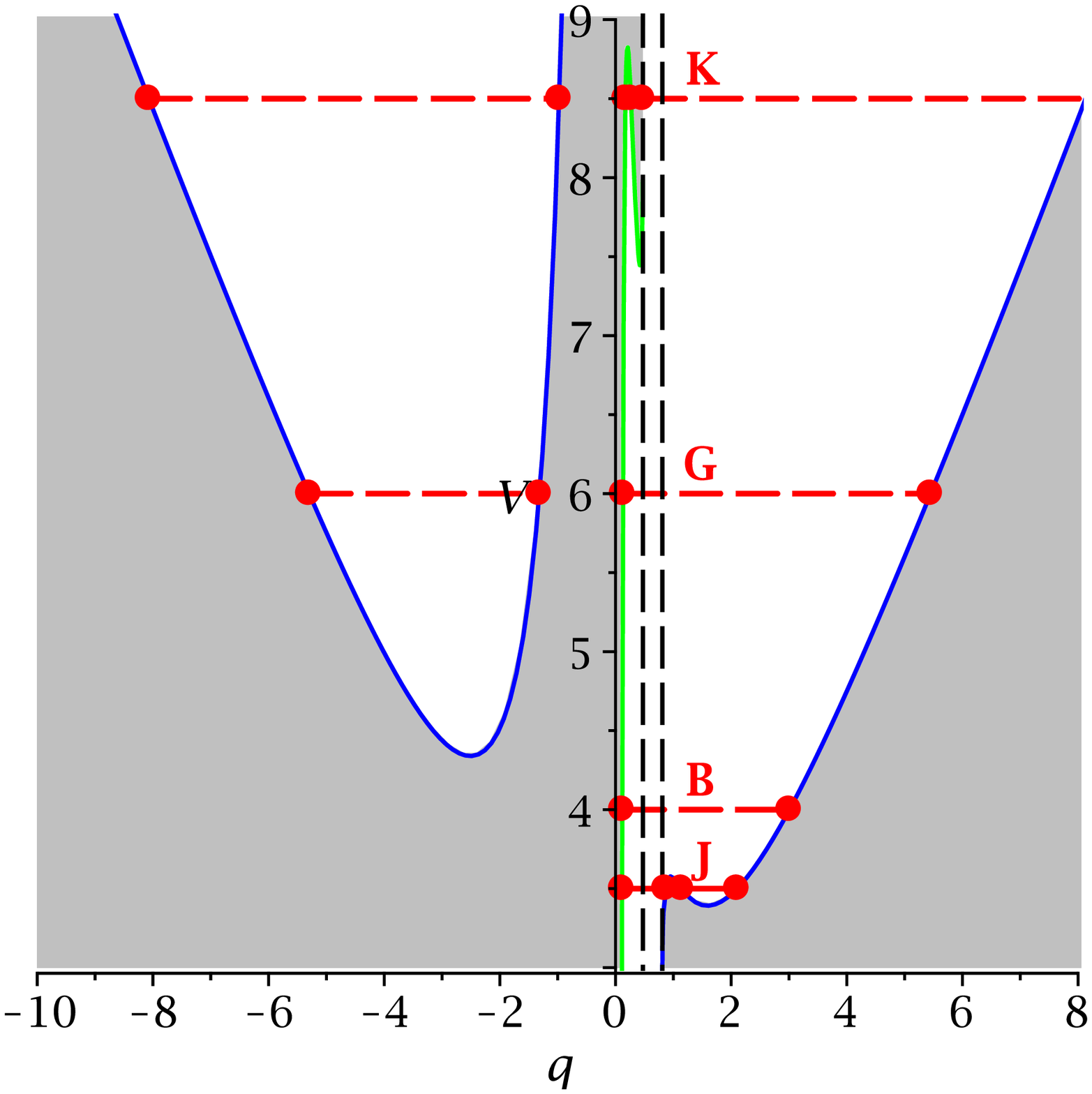}
	}
	\caption{Effective potential $V_{\pm}$ for the $q$-motion. The blue curve represents $V_+$ and the greend curve $V_-$.The dashed vertical lines denote the position of the horizons. The red dashed lines are different energies for different orbit combinations. In the grey area $X(q)<0$, therefore no geodesic motion is possible.}
 \label{pic:potentials}
\end{figure}

We can get more insight into the particle motion by rewriting the polynomial $X$ in the following way
\begin{align}
 X(q) = q^4\left(E-V_+\right)\left(E-V_-\right)\,.
\end{align}
Then we can define the effective potentials $V_{\pm}$
\begin{align}
 V_{\pm} &=-\frac{Ll\pm\sqrt{L(C^2 l^2+l^4-2l^2mq+2l^2q^2+q^4)}}{q^2l} \,,
\end{align}
where now, the zeros of the polynomial $X$ are given by the intersections of the potential $V_{\pm}$ with the energy $E$. Fig. \ref{pic:potentials} shows example plots of the effective potential to visualize the possible orbit combinations given in Tab. \ref{tab:orbit-types}.
To conclude the discussion of the possible orbits, we present a list of all possible orbit types in Tab. \ref{tab:orbit-types}. 
\begin{table}[!ht]
\begin{center}
\begin{tabular}{|lccll|}\hline
type &  zeros & region  & range of $q$ & orbit \\
\hline\hline
A &  0 & I &
\begin{pspicture}(-4,-0.2)(3.5,0.2)
\psline[linewidth=0.5pt]{-}(-4,0)(3.5,0)
\psline[linewidth=0.5pt,doubleline=true](0.3,-0.2)(0.3,0.2)
\psline[linewidth=0.5pt,doubleline=true](1.2,-0.2)(1.2,0.2)
\psline[linewidth=1.2pt]{-}(-4,0)(3.5,0)
\pscircle[hatchcolor=white,fillstyle=solid](-2,0){0.075}
\end{pspicture}
  & TrO
\\  \hline
B & 2 &  II &
\begin{pspicture}(-4,-0.2)(3.5,0.2)
\psline[linewidth=0.5pt]{-}(-4,0)(3.5,0)
\pscircle[hatchcolor=white,fillstyle=solid](-2,0){0.075}
\psline[linewidth=0.5pt,doubleline=true](0.3,-0.2)(0.3,0.2)
\psline[linewidth=0.5pt,doubleline=true](1.2,-0.2)(1.2,0.2)
\psline[linewidth=1.2pt]{*-*}(0,0)(1.5,0)
\end{pspicture}
& MBO
\\  \hline
C   & 2 & III &
\begin{pspicture}(-4,-0.2)(3.5,0.2)
\psline[linewidth=0.5pt]{-}(-4,0)(3.5,0)
\pscircle[hatchcolor=white,fillstyle=solid](-2,0){0.075}
\psline[linewidth=0.5pt,doubleline=true](0.3,-0.2)(0.3,0.2)
\psline[linewidth=0.5pt,doubleline=true](1.2,-0.2)(1.2,0.2)
\psline[linewidth=1.2pt]{-*}(-4,0)(-2.5,0)
\psline[linewidth=1.2pt]{*-}(0,0)(3.5,0)
\end{pspicture}
& EO, TEO 
\\
C$_0$   &  &  &
\begin{pspicture}(-4,-0.2)(3.5,0.2)
\psline[linewidth=0.5pt]{-}(-4,0)(3.5,0)
\pscircle[hatchcolor=white,fillstyle=solid](-2,0){0.075}
\psline[linewidth=0.5pt,doubleline=true](0.3,-0.2)(0.3,0.2)
\psline[linewidth=0.5pt,doubleline=true](1.2,-0.2)(1.2,0.2)
\psline[linewidth=1.2pt]{-*}(-4,0)(-2.5,0)
\psline[linewidth=1.2pt]{*-}(-2,0)(3.5,0)
\end{pspicture}
& EO, TO 
\\ \hline
D & 2 & IV &
\begin{pspicture}(-4,-0.2)(3.5,0.2)
\psline[linewidth=0.5pt]{-}(-4,0)(3.5,0)
\pscircle[hatchcolor=white,fillstyle=solid](-2,0){0.075}
\psline[linewidth=0.5pt,doubleline=true](0.3,-0.2)(0.3,0.2)
\psline[linewidth=0.5pt,doubleline=true](1.2,-0.2)(1.2,0.2)
\psline[linewidth=1.2pt]{*-*}(-2.5,0)(1.5,0)
\end{pspicture}
& CMBO 
\\  \hline
E  & 2 & V &
\begin{pspicture}(-4,-0.2)(3.5,0.2)
\psline[linewidth=0.5pt]{-}(-4,0)(3.5,0)
\pscircle[hatchcolor=white,fillstyle=solid](-2,0){0.075}
\psline[linewidth=0.5pt,doubleline=true](0.3,-0.2)(0.3,0.2)
\psline[linewidth=0.5pt,doubleline=true](1.2,-0.2)(1.2,0.2)
\psline[linewidth=1.2pt]{-*}(-4,0)(-3.5,0)
\psline[linewidth=1.2pt]{*-}(-2.5,0)(3.5,0)
\end{pspicture}
  & EO, CTEO
\\ \hline
F & 4 & VIII  &
\begin{pspicture}(-4,-0.2)(3.5,0.2)
\psline[linewidth=0.5pt]{-}(-4,0)(3.5,0)
\pscircle[hatchcolor=white,fillstyle=solid](-2,0){0.075}
\psline[linewidth=0.5pt,doubleline=true](0.3,-0.2)(0.3,0.2)
\psline[linewidth=0.5pt,doubleline=true](1.2,-0.2)(1.2,0.2)
\psline[linewidth=1.2pt]{*-*}(-2.5,0)(1.5,0)
\psline[linewidth=1.2pt]{*-*}(-3.5,0)(-3,0)
\end{pspicture}
& BO, CMBO
\\ \hline 
G & 4 & IX  &
\begin{pspicture}(-4,-0.2)(3.5,0.2)
\psline[linewidth=0.5pt]{-}(-4,0)(3.5,0)
\pscircle[hatchcolor=white,fillstyle=solid](-2,0){0.075}
\psline[linewidth=0.5pt,doubleline=true](0.3,-0.2)(0.3,0.2)
\psline[linewidth=0.5pt,doubleline=true](1.2,-0.2)(1.2,0.2)
\psline[linewidth=1.2pt]{*-*}(0,0)(1.5,0)
\psline[linewidth=1.2pt]{*-*}(-3.5,0)(-2.5,0)
\end{pspicture}
& BO, MBO
\\ \hline
H  & 4 & X &
\begin{pspicture}(-4,-0.2)(3.5,0.2)
\psline[linewidth=0.5pt]{-}(-4,0)(3.5,0)
\pscircle[hatchcolor=white,fillstyle=solid](-2,0){0.075}
\psline[linewidth=0.5pt,doubleline=true](0.3,-0.2)(0.3,0.2)
\psline[linewidth=0.5pt,doubleline=true](1.2,-0.2)(1.2,0.2)
\psline[linewidth=1.2pt]{-*}(-4,0)(-2.5,0)
\psline[linewidth=1.2pt]{*-*}(0,0)(1.5,0)
\psline[linewidth=1.2pt]{*-}(2,0)(3.5,0)
\end{pspicture}
  & EO, MBO, EO
\\ \hline
I  & 4 & XI &
\begin{pspicture}(-4,-0.2)(3.5,0.2)
\psline[linewidth=0.5pt]{-}(-4,0)(3.5,0)
\pscircle[hatchcolor=white,fillstyle=solid](-2,0){0.075}
\psline[linewidth=0.5pt,doubleline=true](0.3,-0.2)(0.3,0.2)
\psline[linewidth=0.5pt,doubleline=true](1.2,-0.2)(1.2,0.2)
\psline[linewidth=1.2pt]{-*}(-4,0)(-2.5,0)
\psline[linewidth=1.2pt]{*-*}(-1.5,0)(-0.5,0)
\psline[linewidth=1.2pt]{*-}(0,0)(3.5,0)
\end{pspicture}
  & EO, BO, TEO
\\ \hline
J & 4 &  XII &
\begin{pspicture}(-4,-0.2)(3.5,0.2)
\psline[linewidth=0.5pt]{-}(-4,0)(3.5,0)
\pscircle[hatchcolor=white,fillstyle=solid](-2,0){0.075}
\psline[linewidth=0.5pt,doubleline=true](0.3,-0.2)(0.3,0.2)
\psline[linewidth=0.5pt,doubleline=true](1.2,-0.2)(1.2,0.2)
\psline[linewidth=1.2pt]{*-*}(0,0)(1.5,0)
\psline[linewidth=1.2pt]{*-*}(2,0)(3,0)
\end{pspicture}
& MBO, BO 
\\ \hline
K  & 6 & XIII  &
\begin{pspicture}(-4,-0.2)(3.5,0.2)
\psline[linewidth=0.5pt]{-}(-4,0)(3.5,0)
\pscircle[hatchcolor=white,fillstyle=solid](-2,0){0.075}
\psline[linewidth=0.5pt,doubleline=true](0.3,-0.2)(0.3,0.2)
\psline[linewidth=0.5pt,doubleline=true](1.2,-0.2)(1.2,0.2)
\psline[linewidth=1.2pt]{*-*}(0,0)(1.5,0)
\psline[linewidth=1.2pt]{*-*}(-1.5,0)(-0.5,0)
\psline[linewidth=1.2pt]{*-*}(-3.5,0)(-3,0)
\end{pspicture}
  & EO, BO, TEO
\\ \hline
\end{tabular}
\caption{Possible types of orbits for the black spindle spacetime. We represent the range of the orbits by thick lines, whereas the dots mark the turning points of the geodesic motion. $q=0$ is denoted by a single vertical line and the position of the horizons by two vertical lines. }
\label{tab:orbit-types}
\end{center}
\end{table}

\section{Solution of the geodesic equation}\label{sec:sol}
\subsection{Solution for lightlike particles}
If we set $\delta=0$ then $X$ and $Y$ reduce to polynomials of fourth order in $q$ and $p$. Therefore 
\begin{align}
X &= \sum_{i=0}^4a_iq^i \nonumber\\
Y &= \sum_{i=0}^4a'_ip^i \,
\end{align}
with the coefficients
\begin{align*}
a_0 &= Kl^2-L^2& \  
a'_0 &= -K\left(\mathcal{C}^2+l^2\right)+L^2  \\
a_1 &= 0& \  
a'_1 &= 2Km \\
a_2 &= 2EL-2K& \  
a'_2 &= 2EL-2K \\
a_3 &= 0& \  
a'_3 &= 0 \\
a_4 &= -E^2+\frac{K}{l^2}&  
a'_4 &= E^2-\frac{K}{l^2}  \,.
\end{align*}
The substitutions $q=\frac{1}{x}+q_{X}$ and $p=\frac{1}{y}+p_{Y}$ where $q_X$ and $p_Y$ are zeros of $X$ and $Y$, respectively, reduce the order of the polynomials to three
\begin{align}
\left(\frac{\dd x}{\dd\gamma}\right)^2 &=\sum_{i=0}^3b_ix^i \nonumber\\
\left(\frac{\dd y}{\dd\gamma}\right)^2 &=\sum_{i=0}^3b'_iy^i \,,
\label{eqn:poly3}
\end{align} 
with adjusted coefficients $b_i$ and $b'_i$. Further substitutions $x=\frac{1}{b_3}\left(4z-\frac{b_2}{3}\right)$ and $y=\frac{1}{b'_3}\left(4w-\frac{b'_2}{3}\right)$ transform Eqn. \ref{eqn:poly3} into the standard Weierstra{\ss} form
\begin{align}
\left(\frac{\dd z}{\dd\gamma}\right)^2 &=4z^3-g_2z-g_3=P^{q}_3(z) \nonumber\\
\left(\frac{\dd w}{\dd\gamma}\right)^2 &=4w^3-g'_2w-g'_3=P^{p}_3(w) \,.
\label{eqn:standardWeierstrass}
\end{align}
Here $g_2$, $g_3$, $g'_2$ and $g'_3$ denote the Weierstra{\ss} invariants. The solution of Eqn. \ref{eqn:standardWeierstrass} is the elliptic Weierstra{\ss} $\wp$-function \cite{Markushevich:1967}. After resubstitution the solution for $q$ can be written as
\begin{align}
q &=\pm\frac{b_3}{4\wp_{q}\left(\gamma\right)-\frac{b_2}{3}}+q_X \nonumber\\ 
p &=\pm\frac{b'_3}{4\wp_{p}\left(\gamma\right)-\frac{b'_2}{3}}+p_Y \,.
\label{eqn:subs1}
\end{align}
Here 
\begin{align}
 \wp_{q}\left(\gamma\right) &:= \wp\left(\gamma-\gamma_{\rm in};g_2,g_3\right) \nonumber\\
 \wp_{p}\left(\gamma\right) &:= \wp\left(\gamma-\gamma'_{\rm in};g'_2,g'_3\right) \,,
\end{align}
where
\begin{align}
\gamma_{\rm in} &=\gamma_0+\int_{z_0}^{\infty}\frac{\dd z}{\sqrt{4z^3-g_2z-g_3}} \nonumber\\
\gamma'_{\rm in} &=\gamma_0+\int_{w_0}^{\infty}\frac{\dd w}{\sqrt{4w^3-g'_2w-g'_3}} \,
\end{align}
and $\gamma_0$ is the initial value.
Next we will solve the $\sigma$- and $\tau$-equation. Again, for simplicity, we will only solve the first one, since the procedure can be adjusted for the second case. With the help of the $q$- and $p$-equation we can rewrite Eqn. \ref{eqn:EOM} in integral form
\begin{equation}
 \sigma-\sigma_0=\int_{p_0}^p{\frac{L-Ep^2}{\mP}\frac{\dd p}{\sqrt{Y}}}-\int_{q_0}^q{\frac{L+Eq^2}{\mQ}\frac{\dd q}{\sqrt{X}}}=I_p-I_q\,.
\end{equation}
We start with $I_q$. First we apply the same substitution as before to transform $X$ into the Weierstra{\ss} form. A partial fraction decomposition leads to
\begin{align}\label{eqn:eom_parfrac}
 I_q &=\int_{z_0}^z\left(K_0+\sum_{n=1}^4\frac{K_n}{z-p^q_n}\right)\frac{\dd z}{\sqrt{P^q_3(z)}} \\
 I_p &=\int_{w_0}^w\left(K'_0+\sum_{n=1}^2\frac{K'_n}{w-p^p_n}+\sum_{n=1}^2\frac{K''_n}{\left(w-p^p_n\right)^2}\right)\frac{\dd w}{\sqrt{P^p_3(w)}} \,.
\end{align}
Here $p_n$ are first order poles and the $K_n$ are constants arising from the partial fraction decomposition. This elliptic integral of third kind can be solved with the help of the $\wp$, $\zeta$ and $\sigma$ function \cite{Kagramanova:2010bk,Grunau:2010gd,Enolski:2011id} in terms of the Integrals $I_1$ and $I_2$ of reference \cite{Willenborg:2018,Lawden:2008}
\begin{align}
I^{q}_1(v_p) &=\frac{1}{\wp_q'(v_p)}\left[2\zeta_q(v_p)(v-v_p)+\ln{\frac{\sigma_q(v-v_p)}{\sigma_q(v+v_p)}}-\ln{\frac{\sigma_q(v_0-v_p)}{\sigma_q(v_0+v_p)}}\right] \\
I^{q}_2(v_p) &=-\frac{\wp_q''(v_p)}{\wp_q'(v_p)^2}I^q_1-\frac{1}{\wp_q'(v_p)^2}\left[2\wp_q(v_p)(v-v_0)+2\zeta_q(v)+2\zeta_q(v_0)+\frac{\wp_q'(v)}{\wp_q(v)-\wp(v_p)}+\frac{\wp_q'(v_0)}{\wp_q(v_0)-\wp_Q(v_p)}\right]\,,
\end{align}
with $\wp_q'(v_p)=p^q$ and $\zeta_q(z)=-\int{\wp_q(z)}=\frac{\dd}{\dd z}\ln{\sigma_q(z)}$.
With the help of these integrals we can write the solution for $\sigma$ as
\begin{equation}\label{eqn:solsigma}
\sigma-\sigma_0=(K_0-K'_0)(v-v_0)+\sum_{n=1}^4K_nI_1^q(v_{p_n})-\sum_{n=1}^2K'_nI_1^p(v_{p_n})-\sum_{n=1}^2K''_nI_2^p(v_{p_n})\,. 
\end{equation}
\subsection{Solution for massive particles}
For $\delta=1$ the functions $X$ and $Y$ are polynomials of order 6 in $q$ and $p$, respectively. In this section we solve the equation for the coordinate $q$ and the procedure can be easily adjusted for the equation for the coordinate $p$. As a reminder Eqn. \ref{eqn:EOM} can be written as
\begin{equation}
 \left(\frac{\dd q}{\dd\gamma}\right)^2=\sum_{i}^6a_iq^i=P_6(q)\,,
\end{equation}
where the coefficients $a_i$ depend on the parameters of the black hole and of the particle. By substituting 
\begin{equation}
 q=\pm\frac{1}{x}+q_6\,,
\end{equation}
we can reduce the order of the polynomial by $1$. Here $q_6$ is a zero of the polynomial of order $6$. Then we can write
\begin{equation}
 \left(x\frac{\dd x}{\dd\gamma}\right)^2=P_5(x)\,.
\end{equation}
The right hand side of this equation is now a polynomial of order $5$ and the whole equation can be written as a hyperelliptic integral of the first kind
\begin{equation}
 \gamma-\gamma_{\text in}=\int_{x_{\text in}}^x\frac{x'\dd x'}{\sqrt{P_5(x')}} \,.
\end{equation}
By using derivatives of the Kleinian $\sigma$-function $\sigma_i=\frac{\partial \sigma(\vec{z})}{\partial z_i}$ \cite{Hackmann:2008a}, where the $z_i$ are the components of the argument of $\sigma$, we can write the solution as 
\begin{equation}
 x=-\frac{\sigma_1(\vec{\gamma}_{\infty})}{\sigma_2(\vec{\gamma}_{\infty})}\,.
\end{equation}
The vector $\vec{\gamma}_{\infty}$ can be calculated with the help of the initial value $\gamma_{\text in}$ by
\begin{equation}
 \vec{\gamma}_{\infty}=\left(-\int_{x}^{\infty}{\frac{\dd x}{\sqrt{P_5(x)}}}, \gamma - \gamma_{\text in}- \int_{x_{\text in}}^{\infty}{{\frac{x \dd x}{\sqrt{P_5(x)}}}}\right)^T \, .
\end{equation}
A resubstitution leads to the solution of equation \ref{eqn:EOM}
\begin{equation}
 q(\gamma)=\mp\frac{\sigma_2(\vec{\gamma}_{\infty})}{\sigma_1(\vec{\gamma}_{\infty})}+q_6\,.
\end{equation}
Similar to Eqn. \ref{eqn:eom_parfrac} the remaining equations for the coordinates $\sigma$ and $\tau$ include integrals of the form
\begin{equation}
\int_{w_0}^w\frac{1}{\left(w'-p\right)^2}\frac{\dd w'}{\sqrt{P_5(w')}} \,,
\end{equation}
where $P_5(w)$ is a polynomial of order $5$ in $w$. Unfortunately the current methods cannot be used to integrate these terms analytically.

\section{The orbits}\label{sec:orbits}
\begin{figure}[!ht]
\centering
		\subfigure[$3$D]{\includegraphics[width=0.45\textwidth]{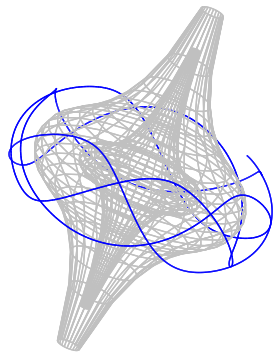}
}
		\subfigure[$2$D]{\includegraphics[width=0.45\textwidth]{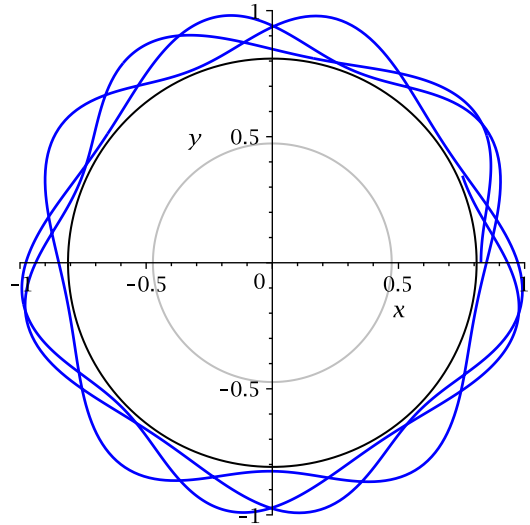}
}
	\caption{Bound orbit for $\delta=0$, $C=0.5$, $l=1$, $m=m_0+0.1$. The blue line represents the orbit. Note that along the line $q=1$ always holds. The grey structures denote the horizons, which are embedded in cylindrical orbits.}
 \label{pic:orbit1}
\end{figure}

\begin{figure}[!ht]
\centering
		\subfigure[$3$D]{\includegraphics[width=0.45\textwidth]{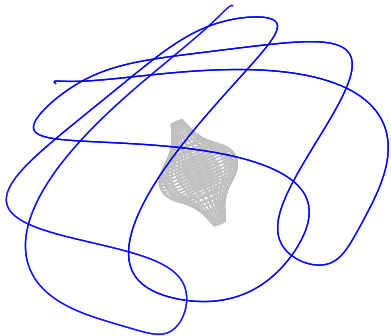}
}
		\subfigure[$2$D]{\includegraphics[width=0.45\textwidth]{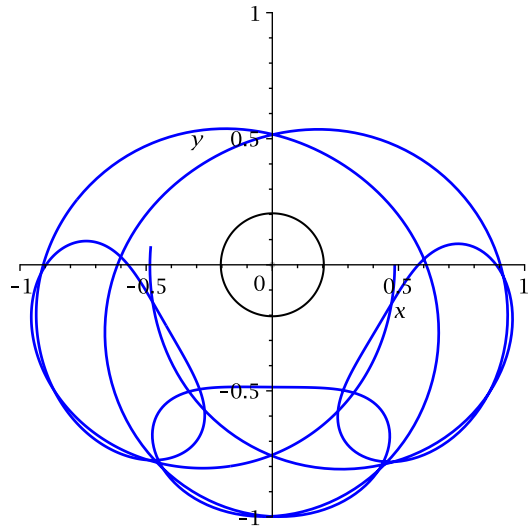}
}
	\caption{Bound orbit for $\delta=0$, $C=0.001$, $l=0.1$, $m=m_0+0.5$. The blue line represents the orbit. Note that along the line $q=1$ always holds. The grey structures denote the horizons, which are embedded in cylindrical orbits.}
 \label{pic:orbit2}
\end{figure}

The orbits we show in this section are visualized in coordinates, where the spindle shape of the spacetime is visible. For the embedding of the horizon, we use the procedure from section \ref{sec:spacetime}. First we set the coordinate $q$ equal to a constant and then we solve Eqn. \ref{eqn:EOM} for one of the constants of motion. Then we embed the orbit to cylindrical coordinates. The resulting orbits are shown in Fig. \ref{pic:orbit1} and Fig. \ref{pic:orbit2}.
\newpage

\newpage
\section{Conclusion}\label{sec:conclusion}
In this article we used the Hamilton-Jacobi formalism to derive the equations of motion for test particles and light for the black spindle spacetime. We analysed the geodesic motion of massive test particles and light and solved the equations of motions in terms of elliptic functions and in part in terms of hyperelliptic functions. The equations of motion for light can be used to calculate observables like the light deflection angle and the shadow of the black spindle.  \\
The analysis could be extended to the case of charged particles and it would be interesting to find a solution for the remaining equations for massive particles.

\section{Acknowledgements}
The authors thank Saskia Grunau and Jutta Kunz  for interesting discussions and remarks. K.F. gratefully acknowledges financial support by the DFG, within the research training group 1620: Models of Gravity.
%------------------ Literatur -----------------%

\bibliographystyle{unsrt}

\end{document}